\documentclass[11pt]{article}
\usepackage{amssymb,cite}
\newcommand\ttl[1]{`#1', }
\newcommand\toline[1]{--#1}
\newcommand{\fract}[2]{{\textstyle\frac{#1}{#2}}}
\newcommand{\ri}{\right}
\newcommand{\ep}{\varepsilon}
\newcommand{\lf}{\left}
\newcommand{\te}{\theta}
\newcommand\eq{\begin{equation}}
\newcommand\en{\end{equation}}
\newcommand\bea{\begin{eqnarray}}
\newcommand\eea{\end{eqnarray}}
\newcommand\nn{\nonumber}

\newcommand\Ai{{\rm Ai}}
\begin{document}
\begin{titlepage}
\vskip 0.5cm
\begin{flushright}
DTP-98/81 \\
ITFA 98-41 \\
{\tt hep-th/9812211} \\
December 1998
\end{flushright}
\vskip 1.5cm
\begin{center}
{\Large{\bf Anharmonic oscillators, the thermodynamic Bethe \\[5pt]
ansatz, and nonlinear integral equations}}
\end{center}
\vskip 0.8cm
\centerline{Patrick Dorey%
\footnote{e-mail: {\tt p.e.dorey@durham.ac.uk}}
and Roberto Tateo%
\footnote{e-mail: {\tt tateo@wins.uva.nl}}}
\vskip 0.9cm
\centerline{${}^1$\sl\small Dept.~of Mathematical Sciences,
University of Durham, Durham DH1 3LE, UK\,}
\vskip 0.2cm
\centerline{${}^2$\sl\small Universiteit 
van Amsterdam, Inst.~voor Theoretische
Fysica, 1018 XE Amsterdam, The Netherlands\,}
\vskip 1.25cm
\begin{abstract}
\noindent
The spectral determinant $D(E)$ of the quartic oscillator is known to 
satisfy a functional equation. This is mapped onto the $A_3$-related 
$Y$-system  emerging in the treatment of a certain perturbed conformal 
field theory, allowing us to give an alternative integral expression 
for $D(E)$. Generalising this result, we conjecture a relationship
between the $x^{2M}$ anharmonic oscillators and the $A_{2M-1}$ TBA
systems. Finally, spectral  determinants for general $|x|^{\alpha}$
potentials are mapped  onto the solutions of nonlinear integral 
equations associated with the (twisted) XXZ and sine-Gordon models. 
\end{abstract}
\end{titlepage}
\setcounter{footnote}{0}
\def\thefootnote{\fnsymbol{footnote}}
%
\noindent
Since the discovery of  quantum mechanics, the 
spectral problem associated with the homogeneous Schr\"odinger operator
\eq
\hat{H} \psi_k(x)= \lf ( - {d^2 \over dx^{2}} + x^{2M} \ri ) \psi_k(x) = E_k 
\psi_k(x)
\label{Sch}
\en
on the real line has been the subject of much attention, with a supply 
of papers which continues to this day:
refs.~\cite{Dun,BW,Ra,HMM,BOW,BPV,V1,V3,V4,FGa} offer 
just a small sample of this work.
Given the apparent simplicity of the system, it is at first
surprising  that
much of the  most  remarkable progress has been made relatively recently.
In the following we will be guided by the theory developed
by Andr\'e Voros in~\cite{V1,V3,V4}, and we refer the reader 
to these articles for a detailed explanation of the subject. 
Here we summarize a few facts that will be needed later.
The confining nature of the potential in (\ref{Sch}) means that the 
spectrum $ \{ E_i \}$ of the theory is discrete. 
The properties of this spectrum can be encoded into 
spectral functions, the simplest example being the spectral determinant
\eq
D_{M}(E)= D_{M}(0) \prod_{k=0}^{\infty} \lf(   1+ {E \over E_k} \ri)
\en
The constant $D_M(0)=\sin(\pi/(2M{+}2))^{-1}$ reflects the
definition of $D_M$ as a zeta-regularized functional 
determinant (see~\cite{V1}).
$D_{M}(E)$ is an entire function of $E$ and the positions of its zeroes
coincide, by definition, with the negated
discrete eigenvalues of eq.~(\ref{Sch}). 
Despite the absence of any closed expression for the $E_k$,
precise  information about
the spectrum can be obtained by various means.
The particular aspect that will be important for us is the fact that the
functions $D_M(E)$ satisfy certain functional equations~\cite{V1,V3},
similar to those previously obtained for related Stokes 
multipliers~\cite{Si}. 
These were
used in~\cite{V1,V3,V4}
to derive sum rules relating the different
eigenvalues, but their utility was limited by the difficulty
in finding solutions to the equations within the class of entire functions.
In this paper we point out a surprising link between these functional
equations and other systems of equations which have arisen in the last few
years in a very different context, namely the
finite-size spectra of integrable 1+1 dimensional quantum field theories.
Numerical work confirms the match, and we feel that 
this unexpected connection between 
two a priori disconnected topics
deserves to be understood at a deeper level.

We begin by reviewing some basic properties of the spectral determinants.
{}From the   Bohr-Sommerfeld approximation  one can deduce
the  asymptotic  positions of the zeroes $E=-E_k\,$:
\eq
b_0 (E_k)^{\mu} \sim 2 \pi ( k + 1/2)~~,\quad k\rightarrow\infty
\label{asy}
\en
where $\mu=(M{+}1)/2M$ and
\eq
b_0 =  \frac{\pi^{1/2}}{M}  
{\Gamma(\fract{1}{2M}) \over \Gamma( \fract{3}{2} +\fract{1}{2M})}~.
\en
In addition, $D_M(E)$ admits a semiclassical asymptotic expansion 
for $|E| \rightarrow \infty$ with $|\arg E|<\pi{-}\delta$, $\delta>0$:
\eq
\ln D_{M}(E) \sim \sum_{j=0}^{\infty} a_j  E^{\mu(1-2j)}
{}~;\quad
a_0= \frac{b_0}{2\sin(\mu \pi)}\,.
\label{detasympt}
\en

Now suppose that $M=2$. In this case  
$D(E) \equiv D_2(E)$ satisfies
the following functional relation~\cite{V1}:
\eq
D(E j^{-1} ) D(E) D(jE)=
D(E j^{-1}) +D(E)+D(j E) +2
\label{vf}
\en
where $j=e^{2 i \pi /3}$. 
Together with the asymptotics just described, this is strongly 
reminiscent of the properties of
solutions to thermodynamic Bethe ansatz (TBA)
equations~\cite{Zam1,KM}.
Consider, for example, the 
perturbation of a theory of
${\mathbb Z}_h$ parafermions by the thermal operator of conformal 
dimensions $\Delta=\bar{\Delta}=2/(h{+}2)$.
This results in an integrable massive quantum field theory,
associated with the $A_{h-1}$ Lie algebra.
There are $h{-}1$ particle species, with masses 
$M_a=M_1\sin(\pi a/h)/\sin(\pi/h)$, $a=1\dots h{-}1$.
Species $a$ and $h{-}a$
are charge-conjugate: $\bar a =h{-}a$.
The scattering theory is factorisable, 
with two particle S-matrix elements~\cite{KS}:
\eq
S_{ab}=\prod_{|a-b|+1 \atop {\rm step~2}}^{{}\atop a+b-1}
\{p\}~~,\qquad a,b=1\dots h{-}1
\en
where, in the notation of~\cite{BCDSa},
$\{p\}=(p{-}1)(p{+}1)\,$,
$(p)={\sinh(\fract{\te}{2}{+} i\fract{\pi p}{2h})%
/\sinh(\fract{\te}{2}{-} i\fract{\pi p}{2h})}$.
Non-perturbative information concerning the finite-size 
scaling functions of the model
in a cylinder geometry can be obtained using the 
thermodynamic Bethe ansatz technique~\cite{Zam1,KM,BLZ4,DT1}.
The simplest instance~\cite{Zam1,KM} expresses
the ground-state energy $E(M_1,R)$ as $-\pi c(M_1R)/6R\,$, where
\eq
c(r)= \frac{3}{\pi^2}
\sum_{a=1}^{h-1} 
\int_{-\infty}^{\infty}\!d\te\,m_ar\cosh\te L_a(\te) \,,
\en 
$L_a(\theta)=\ln(1{+}e^{-\ep_a(\te)})\,$,
$r=M_1R\,$, and $m_a=M_a/M_1$. The functions
$\ep_a(\te)$, $a=1\dots h{-}1$ 
(known as pseudo-energies)
solve the following
equations:
\eq
\ep_a(\te)= m_ar\cosh \te  -  {1 \over 2 \pi}\sum_{b=1}^{h-1} \phi_{ab}*L_a(\te)
\label{tba}
\en 
with $\phi_{ab}(\theta)= - i \partial_{\theta}\ln S_{ab}(\theta)$
and 
$g{*}f(\te)= \int_{-\infty}^{\infty}d\te' g(\te-\te')f(\te')\,$.
Now consider $Y_a(\te)=e^{\ep_a(\te)}$.
These are entire functions of $\theta$, with periodicity 
$Y_a(\te+i\pi(h{+}2)/h)=Y_{\bar a}(\te)$~\cite{Zam2}.
Conjugation symmetry of the ground-state equations means that
$\ep_a(\te)=\ep_{\bar a}(\te)$, so the $Y$'s are entire
functions of $t = \exp({2h}\te/{(h{+}2)})$ on the punctured 
$t$-plane ${\mathbb C}^*={\mathbb C}\setminus \{0\}$.
In fact the $Y$'s are thought (cf.~\cite{DT2})
to be analytic functions of the variables 
$a_{\pm} = (r\exp(\pm\te))^{2h/(h{+}2)}$,
with a finite domain of convergence about the point $(a_+,a_-)=(0,0)$.
The domain  of convergence is finite because of
square-root singularities
linking  the ground state to  excited states~\cite{Zam1,KM,DT1}.
It was also shown in~\cite{Zam2} that the
$Y$'s satisfy a set of functional identities, known as a $Y$-system. 
At $h=4$, $t=\exp(4\theta/3)$, and, taking the conjugation symmetry into
account, the $Y$-system is:
\bea
Y_1(e^{-i\pi/3}t) Y_1(e^{i\pi/3}t) 
&=& 1+Y_2(t)  \label{Y1} \\
Y_2(e^{-i\pi/3}t) Y_2(e^{i\pi/3}t) &=& (1+Y_1(t))^2
\label{Y2}
\eea
Substituting~(\ref{Y1}) into (\ref{Y2}), it is easy to see that
$Y_1(t)$ satisfies a constraint involving itself alone:
\eq
Y_1(e^{-2\pi i/3}t) Y_1(t) Y_1(e^{2\pi i/3}t)=
Y_1(e^{-2\pi i/3}t)+Y_1(t)+Y_1(e^{2\pi i/3}t)+2 \,.
\en
The relation with equation (\ref{vf}) is clear, but
the  analytic properties of $Y_1$  don't quite match 
those of $D$ yet. In particular, $Y_1$ has an essential singularity at
$t=0$. To remedy this, we take a massless limit, replacing the 
driving term $m_ar\cosh \theta$ with $ m_ar e^{\te}$ (this amounts to setting
$a_-=0$).  The resulting $Y$'s are now nonsingular at $t=0$,
and furthermore, for the ground state their zeroes
lie on the line $\Im m\,\te=3\pi/4$, the negative real axis in
the $t$-plane. Setting $m_2r=b_0|_{M{=}2}$ and $t=E$, and identifying 
$Y_1(t)$ with $D(E)$, all of the
standard properties of the spectral determinant of the quartic
oscillator are reproduced.
For example, the large  $\te$ asymptotic  
$Y_2(\te) \sim b_0 e^{\te}$ is
obtained by dropping the convolution term in~(\ref{tba}), and implies
that 
$Y_2(\te)$ takes the value $-1$ at the points 
$\te= x_k+\pi i/2$, where
\eq
b_0 e^{x_k} \sim 2 \pi ( k + 1/2)~~,\quad k\rightarrow\infty\,.
\label{asy1}
\en
Combined with (\ref{Y1}),
this shows
that the zeroes of $Y_1(\te)$  are at $\te=x_k+3\pi i/4$,
matching the asymptotic behaviour~(\ref{asy}). 
Finally, at $t=0$ the solutions of the 
$Y$-system are 
$Y_1=2$, $Y_2=3$,
matching the result $D(0)=2$.
Still unsatisfied,  we performed a numerical 
check.  Eq.~(\ref{tba}) was solved for real $\te$ 
and then, as in~\cite{DT2},
eq.~(\ref{tba}) and the $Y$-system were used to obtain
the values of $Y_1(\te)$ on the line $\Im m\,\te= 3\pi/4$.
The first zeroes were found to high accuracy, and
the resulting predictions for the first five energy levels of the $x^4$
potential are compared with earlier results in Table~I.
%
\begin{table}[htb]
\begin{center}
\begin{tabular}{ c l l } 
\hline 
\hline 
\rule[0.2mm]{0cm}{4.5mm}
{}~k~  &~~~~~$E_k$ (TBA)  &~~~~~~~$E_k$ (QM)  \\[3pt]
\hline
\rule[0.2mm]{0cm}{4mm}
0\,& ~~\,1.06036209048418 & ~~~~\,1.06036209048418289965$^a$      \\
1  & ~~\,3.79967302980139 & ~~~~\,3.79967302980$^b$            \\
2  & ~~\,7.45569793798672 & ~~~~\,7.45569793798673839216$^a$      \\
3  & ~11.64474551137815 &  ~~~11.6447455114$^b$             \\
4  & ~16.26182601885024 &  ~~~16.26182601885022593789$^a$    \\ 
5  & ~21.23837291823595 &  ~~~21.2383729182$^b$      
\rule[-0.2mm]{0cm}{2mm}\\
\hline 
\hline 
\vspace{1pt}
\end{tabular}
\parbox[t]{0.72\linewidth}{\small
Table~I: Energy levels for the $x^4$ potential from the TBA, compared with 
previous results: $a$ from {\protect \cite{BOW,V3}}\thinspace, $b$ from
{\protect \cite{Ra}}} 
\end{center}
\vspace{-0.4cm}
\end{table}

For $M>2$ the equations satisfied by 
$D_M(E)$ become more intricate, and we have yet to 
map them explicitly into known TBA systems.
Instead we took a shortcut,
though later we shall give  an alternative, and more systematic,
treatment of the problem.
The functional relations for  $D_M(E)$
have a ${\mathbb Z}_{h+2}$ symmetry~\cite{V1}, where $h{=}2M$. 
This suggests an examination of $Y$-systems which share
this symmetry in order to find a generalisation of the $M{=}2$ result.
Of the diagonal scattering theories,
this picks out
the models associated with the $A_{h-1}$ or $D_{h/2+1}$ Lie 
algebras (see~\cite{BCDSa,KM,Zam2}), for which the $Y$-systems are
\eq
Y_a(\te - i \fract{\pi}{h})Y_a(\te + i \fract{\pi}{h}) = 
\prod_{b=1}^{r} \lf ( 1+ Y_b(\te) \ri)^{l_{ab}}
\label{ysad}
\en
where $r$ is the rank and $l_{ab}$ the incidence 
matrix of the relevant Dynkin diagram.
However,  the constants $Y_a(\theta{=}{-\infty})$
do not match the value $\sin(\pi/(2M{+}2))^{-1}$
of $D_M(0)$.
But all is not lost: we can invoke another system of functional
relations, related to the $Y$-system, called the $T$-system 
(cf.~\cite{BR}):
\eq
T_a(\te - i \fract{\pi}{h})T_a(\te + i \fract{\pi}{h}) = 
1+\prod_{b=1}^{r}  T_b(\te)^{l_{ab}} ,
\en
with $Y_a(\theta) =\prod_{b=1}^{r}  T_b(\te)^{l_{ab}}$. When
$M{=}2$ we have $T_2(\te)=Y_1(\te)$, and so we can also
search for our generalisation amongst the $T$-systems.
Asymptotic checks lead to the conjecture that
$D_M(E)$ coincides with the function $T_{M}(\theta)$ of the
massless $A_{2M-1}$ TBA system
obtained from eq.~(\ref{tba}) by setting $h{=}2M$, replacing
the terms $m_ar\cosh \te$ 
by $m_are^\te$, and setting  $m_Mr=b_0$
and  $e^{\te/\mu}=E$.
This was checked using the fact that the zeroes of $T_M(\te)$ on the line
$\Im m\,\te=(h{+}2)\pi/2h$ correspond to zeroes of $1{+}Y_M(\te)$ on the
line $\Im m\,\te=\pi/2$, and these can be located using (\ref{tba}) and the
$Y$-system as before.
Tables~II and~III show some results
for $M=3$ and~$4$.

%
\begin{table}[htb]
\begin{center}
\begin{tabular}{ c l l } 
\hline
\hline
\rule[0.2mm]{0cm}{4.5mm}
{}~k~  & ~~~~~$E_k$ (TBA)  & ~~~~~~~$E_k$ (QM)  \\[3pt]
\hline
\rule[0.2cm]{0cm}{2mm}
0\,  &  ~~\,1.144802453797075~  &  ~~~~\,1.14480245379707$^a$ \\
1  &  ~~\,4.338598711513990  &   ~~~~\,4.3385987115$^b$       \\
2  &  ~~\,9.073084560921449  &   ~~~~\,9.07309$^c$        \\
3  &  ~14.93516963491078   &   ~~~14.9351696349$^b$
\rule[-0.2mm]{0cm}{2mm}\\ 
\hline
\hline
\vspace{1pt}
\end{tabular}
\parbox[t]{0.72\linewidth}{\small
Table~II: Energy levels for the $x^6$ potential from the TBA 
compared with previous results:
$a$ from table~2 of {\protect \cite{FGa}}, 
$b$ from table~I of {\protect \cite{V3}}, and $c$ from
table VII of  {\protect \cite{HMM} }, rescaled by $2^{3/4}$}
\end{center}
\vspace{-0.2cm}
\end{table}
%
%
\begin{table}[htb]
\begin{center}
\begin{tabular}{ c l l } 
\hline
\hline
\rule[0.2mm]{0cm}{4.5mm}
{}~k~  & ~~~~~$E_k$ (TBA)  & ~~~~~~~$E_k$ (QM)  \\[3pt]
\hline   
\rule[0.2cm]{0cm}{2mm}
0\,  &   ~~\,1.2258201138005~ &  ~~~~\,1.22582011382$^a$~ \\
1  &   ~~\,4.7558744139607 &   ~~~~\,4.7558$^c$    \\
2  &   ~10.2449469772369 &   ~~~10.2450$^c$       \\
3  &   ~17.3430879705857 &   ~~~17.3433$^c$
\rule[-0.2mm]{0cm}{2mm}\\ 
\hline   
\hline
\vspace{1pt}
\end{tabular}
\parbox[t]{0.72\linewidth}{\small
Table~III: Energy levels for the $x^8$ potential from the TBA 
compared with previous results:
$a$ from table~2 of {\protect \cite{FGa}}, 
and $c$ from
table VII of {\protect \cite{HMM}}, rescaled by $2^{4/5}$ } 
\end{center}
\vspace{-0.4cm}
\end{table}

The story might have ended here, but in fact it goes
considerably
further. Following~\cite{V3}, we begin by asking about potentials of 
odd degree,
so that the confining potential is $|x|^{2M}$, with $M$ now allowed to be a
half-integer. It helps to split the eigenvalues according to the parity of 
their
eigenfunctions, decomposing $D(E)$ accordingly as
$D(E)=D^+(E)D^-(E)$, with
\eq
D^{\pm}(E)
= D^{\pm}(0) \prod_{k {{\rm even}\atop{\rm odd}}} 
\lf( 1+{E \over E_k}\ri).
\en
These spectral subdeterminants together satisfy a rather simpler equation 
than that obeyed by the full spectral determinant, which also holds if $M$
is a half-integer~\cite{V1}:
\eq 
\Omega^{1/2}D^+(\Omega^{-1}\!E)D^-(\Omega E)
-\Omega^{-1/2}D^+(\Omega E)D^-(\Omega^{-1}\!E)=2i
\label{D+D-}
\en
where $\Omega=e^{i\pi/(M{+}1)}$. This is very similar to the
so-called `quantum Wronskian' condition satisfied by the ${\bf Q}$-operators
introduced in ref.\cite{BLZ2}. The 
similarity becomes more striking when the condition is written in terms
of the operators ${\bf A}_{\pm}(\lambda)\equiv 
\lambda^{\mp 2P/\beta^2}{\bf Q}_{\pm}(\lambda)\,$:
\eq
q^{2P/\beta^2}{\bf A}_+(q^{1/2}\lambda){\bf A}_-(q^{-1/2}\lambda) 
- q^{-2P/\beta^2}{\bf A}_+(q^{-1/2}\lambda){\bf A}_-(q^{1/2}\lambda)
= 2i\sin(2\pi P)\,.
\en
The parameters $q$ and $\beta^2$ are related by
$q=e^{i\pi\beta^2}$.
As it stands, this is an operator equation, but it becomes a functional 
equation when applied to the simultaneous
eigenvectors $|\alpha,p\rangle$ of the operators ${\bf A}_{\pm}(\lambda)$ and 
$P$, 
defined via 
${\bf A}_{\pm}(\lambda)|\alpha,p\rangle=A_{\pm}(\lambda,p)|\alpha,p\rangle$,
$P|\alpha,p\rangle=p|\alpha,p\rangle$.
(We refer the reader to~\cite{BLZ2} for the background to these
definitions.) For brevity we will leave the $p$-dependence of the eigenvalues
$A_{\pm}(\lambda,p)$ implicit from now on;
they are entire functions of
$\lambda^2$, with $A_{\pm}(0)=1$, and a finite number of complex and negative
real zeroes. The remaining zeroes accumulate towards $+\infty$ along the
positive real axis of the $\lambda^2$-plane.
To choose particular eigenvalues
as candidate spectral subdeterminants, we recall
that the zeroes of $D^{\pm}(E)$ lie on the negative real $E$ axis.
This selects the `vacuum eigenvalues' $A^{(v)}_{\mp}(\lambda)$, for
which {\sl all}\ of the zeroes lie on the real axis of the $\lambda^2$-plane, 
and
suggests to identify
$A_{\mp}^{(v)}(\nu E^{1/2})$ with $\alpha^{\pm}D^{\pm}(-E)$, using
the following dictionary, where as before $\mu=(M{+}1)/2M\,$:
\bea
&&\beta^2= 1/(M{+}1)~,\quad
p=1/(4M{+}4)\nn\\[3pt]
&&\nu=(2M{+}2)^{-1/2\mu}\,\Gamma\bigl(\fract{1}{2\mu}\bigr)^{-1}\nn\\[3pt]
&&\alpha^{\pm}= \sqrt{\pi}\,(2M{+}2)^{\mp 1/4\mu}\,%
\Gamma\bigl(\fract{1}{2}{\pm}\fract{1}{4\mu}\bigr)^{-1} 
\label{dict}
\eea
Note, $\alpha^+\alpha^-= \sin\pi/(2M{+}2)$.
The constant~$\nu$ is fixed by comparing the behaviour of
$A_{\mp}(\lambda)$ as $\lambda^2\rightarrow -\infty$~\cite{BLZ2} with that of
$D^{\pm}(E)$ as $E\rightarrow +\infty$~\cite{V4}:
\bea
\ln A_{\mp}(\lambda)\,&\sim &(M{+}1)\Gamma\bigl(\fract{1}{2\mu}\bigr)^{2\mu}%
a_0\,(-\lambda^2)^{\mu}~; \nn\\[3pt]
\ln D^{\pm}(E)&\sim & \fract{1}{2}\,a_0\,E^{\mu}
\qquad~~~
(\,a_0=b_0/(2\sin\mu\pi)\,)\,.
\label{asmyptcomp}
\eea
Finally, the zeroes of $A_{\mp}^{(v)}(\lambda)$ should all
lie on the {\it positive}\ real axis of the $\lambda^2$-plane if 
they are to map onto those of $D^{\pm}(E)$. This holds if
$\mp 2p>-\beta^2$~\cite{BLZ2}, a condition which is indeed met here.
For a more precise check, we sought some numerical evidence. 
As in~\cite{BLZ2}, consider the functions 
$a^{(v)}_{\pm}(\lambda)=e^{\pm 4\pi
ip}A^{(v)}_{\pm}(q\lambda)/A^{(v)}_{\pm}(q^{-1}\lambda)$.
The so-called T-Q relation implies that they
assume the value $-1$ precisely at the zeroes either of
$A^{(v)}_{\pm}(\lambda)$, or of a related entire function $T(\lambda)$. 
For the vacuum eigenvalues, the zeroes of $T(\lambda)$ are away from the
positive real axis and so a search of this line for zeroes of
$a^{(v)}_{\pm}(\lambda){+}1$ will allow us to locate the zeroes of 
$A^{(v)}_{\pm}(\lambda)$.
At the values of $p$ and $\beta$ given by (\ref{dict}),
the functions 
$f_{\pm}(\te)\equiv\ln a^{(v)}_{\pm}(e^{\te/2\mu})$ 
solve 
the following nonlinear integral equations (NLIE):
\bea
f_{\pm}(\theta)&=&
-\fract{1}{2}i\,b_0\nu^{-2\mu}e^{\theta}
+\int_{{\cal 
C}_1}\!\varphi(\theta{-}\theta')\ln(1{+}e^{f_{\pm}(\theta')})\,d\theta'
\nn\\[2pt]
&&\qquad\qquad\qquad{}-
\int_{{\cal 
C}_2}\!\varphi(\theta{-}\theta')\ln(1{+}e^{-f_{\pm}(\theta')})\,d\theta'
\pm i\pi/2 
\label{nlie}
\eea
where the contours ${\cal C}_1$
and ${\cal C}_2$ run from $-\infty$ to $+\infty$, just below and just
above the real $\theta$-axis, and
\eq
\varphi(\theta)=\int_{-\infty}^{\infty}%
\frac{e^{i\omega\theta}\sinh\fract{\pi}{2}(\xi{-}1)\omega}%
{2\cosh\fract{\pi}{2}\omega\sinh\fract{\pi}{2}\xi\omega}%
\frac{d\omega}{2\pi}~~~,\qquad \xi=\fract{1}{M}\,.
\label{krnl}
\en
Such equations first arose in
refs.~\cite{KBP,DDV}, in the contexts of the (twisted) XXZ model,
and the sine-Gordon model at coupling $\beta$.

Solving (\ref{nlie}) numerically, we can now test
the conjecture (\ref{dict}). For $M=2,3,4$,
the results of Tables~I,II,III were reproduced, with 
disagreements being typically in the last quoted digit of the TBA data.
Next we set $M=3/2$, and obtained the results quoted in~\cite{V4} 
for the $|x|^3$ potential.  It was then natural to conjecture that the
identification remains valid at arbitrary $M>1$. In the absence of
suitable published data, we used the {\sc Maple} package to diagonalise 
the Hamiltonian~(\ref{Sch}) in a basis of 
harmonic oscillator eigenfunctions, as in~\cite{V4}. Agreement 
with (\ref{nlie}) was confirmed for various potentials $|x|^{2M}$;
some results for $M=15/8$ are shown in Table~IV.

\vspace{7pt}
\begin{table}[htb]
\begin{center}
\begin{tabular}{ c l l } 
\hline
\hline
\rule[0.2mm]{0cm}{4.5mm}
{}~k~  & ~~~~~$E_k$ (NLIE)  & ~~~~~~~$E_k$ (QM)  \\[3pt]
\hline
\rule[0.2cm]{0cm}{2mm}
0\,  & ~~  1.0503451122723      &  ~~~~~ 1.05034511~        \\ 
1  &  ~~ 3.7190710425856      &  ~~~~~ 3.71907104      \\
2  &  ~~ 7.2061514537816      &  ~~~~~ 7.20615145      \\
3  &  ~~ 11.148641889036      &  ~~~~~ 11.1486419
\rule[-0.2mm]{0cm}{2mm} \\ 
\hline
\hline
\vspace{1pt}
\end{tabular}
\parbox[t]{0.72\linewidth}{\small
Table~IV:
Energy levels for the $|x|^{15/4}$ potential computed using eq.~(\ref{nlie}),
compared with direct QM results.} 
\end{center}
\vspace{-0.4cm}
\end{table}

For $M\le 1$, the formulae for the determinants become divergent and 
need further regularisation~\cite{V4}; at the same time, the calculations
of~\cite{BLZ2} depart from the so-called `semiclassical domain' and
must be modified.
Nevertheless, we have evidence that the correspondence
continues to hold. At $M{=}1$,
the sine-Gordon model is at the free-fermion point, the
kernel~(\ref{krnl}) vanishes, and
the energy levels $E_k=(2k{+}1)$ of the simple harmonic oscillator
are easily recovered. Then at $M{=}1/2$, the $D^{\pm}(E)$ are known in closed
form~\cite{V4}, leading to the
predictions
\bea
a^{(v)}_+(\nu E^{1/2})\,%
&=&\,~\Omega\,\Ai(-\Omega^2E)/\Ai(-\Omega^{-2}E)\,,\nn\\
a^{(v)}_-(\nu E^{1/2})\,%
&=&\,\Omega^{-1}\Ai'(-\Omega^2E)/\Ai\,'(-\Omega^{-2}E)\,,
\eea
where $\Ai(E)$ is the Airy function and $\Omega=e^{2\pi i/3}$.
These were verified to 15 digits. Note that $\beta^2{=}2/3$ for $M{=}1/2$:
this is the $N{=}2$ supersymmetric point of the sine-Gordon model, and
it is tempting to conjecture 
a link with the Painlev\'e III results of ref.~\cite{FS}, though this
remains to be elucidated. Finally, we made a numerical check against 
{\sc Maple} results at $M{=}7/8$, again finding agreement.

When $M$ is an integer the potential is analytic; it is interesting that these
cases are mapped onto the reflectionless points of the sine-Gordon model.
In the TBA framework, these are described by 
$D_{2M-1}$-related systems, with the twist $p=1/(4M{+}4)$ implemented by
introducing fugacities $\pm i$ on the fork nodes $2M{-}2$ and $2M{-}1$
(see~\cite{FS} for similar man\oe uvres in the repulsive regime).
It can then be checked that, for the ground state with
$\ep_{2M-2}=\ep_{2M-1}$, this is equivalent to an $A_{2M-1}$-related
system, thus making a link with the approach described in the first half
of this paper.

\vspace{5pt}
\noindent{\bf Acknowledgements -- }
We are grateful to Ferdinando Gliozzi, Bernard Nienhuis
and especially Andr\'e Voros for useful discussions.
The work was supported in part by a TMR grant of the
European Commission, reference ERBFMRXCT960012.
PED  thanks the EPSRC for an Advanced Fellowship, and
RT thanks  SPhT Saclay for hospitality
and the Universiteit van Amsterdam for a post-doctoral fellowship.
%

\end{document}